\documentclass[epj]{svjour}
\usepackage{graphics}
\begin{document}
\title{Optical Current, Momentum and Angular Momentum in Anisotropic Materials exposed to Detailed Balance}
\author{B.A. van Tiggelen\inst{1} 
}                     
%
%
\institute{Univ. Grenoble Alpes, CNRS, LPMMC, 38000 Grenoble, France}
\date{Received: date / Revised version: date}
%
\abstract{This work investigates the theory behind a thought  experiment that intends to measure momentum and angular momentum of matter exposed to ``isotropic radiative noise". Radiative momentum has been a controversial subject for decades. The momentum of isotropic noise is intuitively expected to be zero.
We formulate the general features of the isotropic noise such as equipartition of energy and the  vanishing of integrated Poynting vector. We demonstrate that in bi-anisotropic materials, a finite radiative momentum persists that performs work on the matter when its parameters change slowly. We find that  Faraday rotation in the scattering of the radiative noise induces an angular momentum along the applied external magnetic field. Also, a Poynting vector starts circulating around the matter, raising the question whether it really describes the energy flow. These effects are small and hard to measure in any real experiment. Yet, they are surprising predictions of the classical, macroscopic Maxwell equations, and make contact with the outcome of recent QED calculations done for the quantum vacuum.
%
} 
\titlerunning{optical current, momentum and angular momentum in noise}
\maketitle

\section{Introduction}
Generally speaking, ``detailed balance" refers to a a situation where each elementary process is equilibrated by its reverse process. Applied to light propagation in matter, it means that absorbtion is compensated by emission, and that scattering of light from $a$ to $b$ is compensated by the scattering from $b$ to $a$. Detailed balance is typically realized at all frequencies in thermodynamic equilibrium, in particular the ''quantum vacuum" at zero temperature can be said to be in detailed balance. Equipartition and isotropy also occur in the presence of perfect multiple light scattering. In this case, the energy of the radiation is equipartitionned among all micro-states and no macroscopic energy current exists in the medium. Two subtle complications show up in more advanced media. In an anisotropic medium, energy current and momentum are not necessarily equal, and they cannot vanish simultaneously. Secondly, a Poynting vector $\mathbf{S}$ that vanishes everywhere looks like a too strong statement to be true for an inhomogeneous medium. If indeed it does not vanish everywhere, it can integrate to zero yet produce   a finite angular momentum $\int d^3 \mathbf{r}\, \mathbf{r} \times \mathbf{S}$. 

In this work we discuss the possibility to create a non-zero momentum density of the radiation generated by the detailed balance. We are aware of the one-century-old controversy on how to define the momentum of radiation in matter \cite{PRBrevik,Robinson,peierls,nelson,loudon}, with recent discussions for spin and orbital angular momentum \cite{Bliokh1}. In this work we will adopt $\mathbf{E}\times \mathbf{B}/4\pi c_0$ as the momentum density of light (the so-called Nelson version \cite{nelson}). This vector density obeys a conservation law that is not \emph{explicitly} affected by constitutive assumptions, identical to its microscopic version, with symmetric 3D stress tensor. Imposing current and momentum to coincide up to a factor $c_0^2$ has been one of the main arguments in favor of the so-called Abraham version $\mathbf{E}\times \mathbf{H}/4\pi c_0$ as it would symmetrize momentum and energy in the 4D stress tensor \cite{jackson}. A link has been established with the difference between ``canonical" and ``kinetic" momentum density \cite{barnett}, that differ precisely by the curl of a vector field \cite{Bliokh2} so that both total momenta should be equal. In the case of an isotropic radiation field discussed here, the ``canonical"  momentum density is shown to vanish, unlike the kinetic momentum density. We will  predict a non-zero radiation (angular) momentum in the presence of isotropic electromagnetic noise that, unfortunately, will be hard to test  experimentally.

\section{Field correlations for detailed balance}

In the following we consider the presence of propagating electromagnetic fields in a conservative medium generated by random sources.
In the simplest classical picture, the interaction of electromagnetic fields with matter is described by constitutive equations, with neglect of high multi-poles, and slowly varying external fields that enter the constitutive equations adiabatically. We usually write them as

\begin{eqnarray}\label{CE}
 \nonumber \mathbf{ D }&=& \varepsilon (\omega) \cdot \mathbf{E} + \chi (\omega) \cdot \mathbf{B }\\
          \mathbf{ H}  &=& - \chi^*(\omega) \cdot \mathbf{E} + \mu^{-1} (\omega)\mathbf{B }
\end{eqnarray}
featuring 3 material tensors $\mu$, $\varepsilon$ and $\chi$.  To guarantee conservation of energy we will assume $\varepsilon$ and $\mu$ to be  hermitian operators ($\varepsilon =\varepsilon^\dag$) that may exhibit dynamic ($\omega$) and spatial  dispersion, and may also be inhomogeneous. So in general we write $\varepsilon  = \varepsilon (\omega, \mathbf{r},\mathbf{p})$ with $\mathbf{p}$ the hermitian operator $-i\partial_\mathbf{r}$ with eigenvalue $\mathbf{k}$ in a homogeneous medium and the position operator $\mathbf{r}$ with eigenvalue $\mathbf{x}$. Since frequency dependence originates from time-convolution of real-valued (observable) operators, all material parameters must be hermitian functions of $i\omega$  and be analytic in the upper complex frequency plane to prevent non-causal behavior. This means that Faraday rotation - described by a dielectric tensor $\varepsilon_{nm} (\omega) = V\epsilon_{nmk}B^0_k c_0 / i\omega $ - and rotatory power - with bi-anisotropic tensor $\chi_{nm} =  g\delta_{nm} c_0/i\omega$ -  are covered by the present approach.  A  dissipative part will be added below, assumed small, since energy current density and energy current are in principle only well-defined in the limit of vanishing dissipation. In the simplest form of bi-anisotropy, one ignores spatial dispersion in $\varepsilon$, $\mu$ and $\chi$, knowing that some spatial dispersion, such as the rotatory power mentioned earlier, is already generated by the presence of  $\chi$.

Finally, we will be  interested in slow changes (compared to the cycle time $1/\omega$) in the constitutive parameters, for instance induced by a slow change in an externally applied magnetic field $\mathbf{B}_0(t)$. The slowness guarantees the adiabatic following of steady-state solutions. We will show that such time-dependence preserves momentum conservation of matter and radiation.

\subsection{Fluctuation-Dissipation}
In the following we will consider a situation of \emph{stationary detailed balance} over a broad range of frequencies.
To achieve a stationary balance, equals amounts of energy must be injected and removed. The fundamental relation between source and dissipation is known as the fluctuation-dissipation theorem \cite{FD}.  We will adopt a current density $\mathbf{J}= \mathbf{J}_0 + \mathbf{j}$, with $\mathbf{J}_0$ the \emph{electric} current density induced by the electric field, described by a  local Ohm's law $\mathbf{J}_0(\mathbf{x},\omega) = \sigma(\mathbf{x},\omega) \cdot \mathbf{E}(\mathbf{x},\omega)$ which will generate dissipation rate proportional to the local conductivity tensor at optical frequency $\sigma(\mathbf{x},\omega)$, and $\mathbf{j}$  a random ''noise" current density that will act as a source in Maxwell's equations. For simplicity, we will assume  the noise to be stationary and uncorrelated, but with the possibility to be anisotropic and inhomogeneous,

\begin{equation}\label{fd}
  \langle j_j(\mathbf{x},\omega)\overline{j}_l(\mathbf{x}',\omega')\rangle  = 2Q(\omega)\sigma_{jl}(\omega,\mathbf{x}) \delta(\mathbf{x}-\mathbf{x}')  \delta_{\omega\omega'}
\end{equation}
with the notation $\delta_{\omega\omega'}= 2\pi \delta(\omega-\omega')$ to get rid of factors of $2\pi$.
The fluctuation-dissipation formula relates local electric current fluctuations to the local AC electrical conductivity tensor $\sigma_{jl} (\omega,\mathbf{x})$. The latter is hermitian and has (in Gaussian units) the dimension of inverse time since $\mathbf{J}$ has the dimension of $\partial_t \mathbf{E}$). It is important for what follows that the energy $Q(\omega)$ be independent on $\mathbf{x}$, expressing local detailed balance. If it is thermal, $Q(\omega) = \hbar \omega$ $\times \left( 1 - \exp(-\hbar \omega/kT)\right)^{-1}$ both for negative and positive frequencies (implying, quite subtly, that at  $T=0$ only positive frequencies come in).

In the presence of a finite AC conductivity tensor, the retarded (complex frequency $\omega + i\epsilon$) Green's operator associated with the  Helmholtz equation for the electric field is

\begin{eqnarray}\label{Gr}
  \nonumber &&\mathbf{G}(\mathbf{r},\mathbf{p},\omega) = \\
 \nonumber && \left[ \frac{\omega^2}{c_0^2} \mathbf{\mathcal{E}(\mathbf{r})} + \mathbf{L}(\mathbf{r},\mathbf{p})^\dag \cdot \mu(\mathbf{r})^{-1} \cdot \mathbf{L}(\mathbf{r},\mathbf{p})  + \frac{4\pi i \omega }{c_0^2}\mathcal{\sigma}(\mathbf{r})  \right]^{-1}\\
\end{eqnarray}
in terms of the hermitian  operator  $\mathbf{\mathcal{E}}= \varepsilon+ \chi\cdot \mu \cdot \chi^\dag$ and the operator $\mathbf{L}(\mathbf{p},\mathbf{r})= \epsilon_\mathbf{p} - (\omega/c_0) \mu(\mathbf{r}) \cdot \chi(\mathbf{r})^\dag$. We have dropped explicit reference to frequency dependence of material parameters. Note that $\mathbf{G}^\dag(\sigma)=\mathbf{G}(-\sigma)$ since all individual ingredients other than $i\omega\times \sigma $ in   $\mathbf{G}$  are hermitian. We can thus write down the formal solution of the Helmholtz equation with $4\pi \partial_t \mathbf{j}/c^2_0$ as source as $E_i(\mathbf{x},\omega)= \int d^3\mathbf{x}' G_{ik}(\mathbf{x}, \mathbf{x}',\omega) (4\pi i\omega/c_0)  j_k(\mathbf{x}',\omega)$. Since the source is proportional to frequency, we denote $\mathbf{G}^s=\omega \mathbf{G}$, so that

\begin{eqnarray}\label{GG}
  \nonumber    && \langle E_i(\mathbf{x}_1,\omega)\overline{E}_j(\mathbf{x}_2,\omega')\rangle  \\
 \nonumber   &&= \left(\frac{4\pi}{c_0^2}\right)^2 2Q \delta_{\omega\omega'}  \int d^3\mathbf{x}' \int d^3\mathbf{x}'' \\
 \nonumber && \ \ \ \ \ \ \ \  G^s_{ik}(\mathbf{x}_1, \mathbf{x}',\omega)   \overline{G}^s_{jl}(\mathbf{x}_2, \mathbf{x}'',\omega) \sigma_{kl}(\mathbf{x}') \delta(\mathbf{x}'-\mathbf{x}'')\\
  \nonumber   && = \left(\frac{4\pi}{c_0^2}\right)^2 2Q \delta_{\omega,\omega'}   \langle \mathbf{x}_1| G^s_{ik}(\mathbf{r},\mathbf{p},\omega)  \sigma_{kl}(\mathbf{r})  G^{s\dag}_{lj}(\mathbf{r},\mathbf{p},\omega)  |\mathbf{x}_2\rangle \\
\nonumber  &&= \frac{4\pi \omega Q }{c_0^2 i} \delta_{\omega\omega'} \left[ G^\dag_{ij} (\mathbf{x}_1,\mathbf{x}_2,\omega ) - G_{ij} (\mathbf{x}_1,\mathbf{x}_2,\omega ) \right]\\
  &\, &
\end{eqnarray}
The transposition operation $\dag$ is both with respect to polarization and to space coordinates. In the last equality we used the operator identity $ \mathbf{G}^\dag-\mathbf{G }= \mathbf{G} \cdot(\mathbf{G}^{-1} - \mathbf{G}^{\dag,-1})\cdot \mathbf{G}^\dag$. This is nothing but one version of the celebrated ''$\mathrm{Im}\, G$ theorem"  and is widely used to measure "passively'' the Green's function by cross-correlating stationary noise \cite{passive1,passive2,passive3,bartprl}. It is valid under very broad conditions, and applies when all parameters are anisotropic and inhomogeneous, even for finite dissipation, the only constraint being that $Q(\omega)$ be homogeneous. It also applies to a stationary multiple scattering process where a steady energy flux density $F(\omega)$ of an external source is compensated by leaks through the medium boundaries. In that case is $Q(\omega) \approx F(\omega)c_0^2/\omega^2$ \cite{prlbart}, which can be much larger than the typical thermal value $kT$ at the expense of a much smaller bandwidth.

\subsection{Energy and Current in homogeneous media}

For homogeneous media with detailed balance the analysis simplifies, and allows us to identify energy density and energy current in the presence of dispersion. We define the Wigner function (no confusion exists between the operator $\mathbf{r}$ and its eigenvalue $\mathbf{r}$ or $\mathbf{x}$),

\begin{eqnarray}\label{WF}
 \nonumber   \phi_{ik,\mathbf{k}}(\Omega,  \mathbf{r}) &=&  \int d^3\mathbf{x} e^{-i\mathbf{k}\cdot \mathbf{x}} E_i(\omega_+,\mathbf{r}+\frac{1}{2}\mathbf{x}) \overline{E}_k(\omega_-,\mathbf{r}-\frac{1}{2}\mathbf{x}) \\
\end{eqnarray}
where $\omega_\pm = \omega \pm \frac{1}{2}\Omega$. Its Fourier transform is $\phi_{ik,\mathbf{k}}(\Omega, \mathbf{q})$,  $\phi_{ik,\mathbf{k}}(\Omega, \mathbf{q})d^3\mathbf{k}d\Omega d\omega$ being an energy.  Since
\begin{eqnarray*}
e^{-i\mathbf{k}\cdot \mathbf{x}} e^{-i\mathbf{q}\cdot \mathbf{r}}
= e^{-i(\mathbf{k}+\frac{1}{2}\mathbf{q})\cdot(\mathbf{r}+\frac{1}{2} \mathbf{x})}
  e^{i(\mathbf{k}-\frac{1}{2}\mathbf{q})\cdot(\mathbf{r}-\frac{1}{2} \mathbf{x})}
\end{eqnarray*}
we see that $\phi_{ik,\mathbf{k}}(\Omega,   \mathbf{q}) = \langle E_i(\omega_+,\mathbf{k}_\pm) \overline{E}_k(\omega_-,\mathbf{k}_-) \rangle$ with $\mathbf{k}_\pm = \mathbf{k} \pm \frac{1}{2}\mathbf{q}$, when averaged over the random source. Let us further define
\begin{eqnarray*}
&& \mathcal{W}_\pm(\Omega,\mathbf{q},\sigma)= \\
&&\frac{\omega_\pm }{c_0^2} \mathcal{E}(\omega_\pm) + \omega_\pm^{-1}\mathbf{L}(\mathbf{k}_\pm,\omega_\pm,)^\dag \cdot \mu(\omega_\pm)^{-1} \cdot \mathbf{L}(\mathbf{k}_\pm,\omega_\pm)    \\
&& \ \ \ \ \ \ \ \pm \frac{4\pi i\sigma(\omega_\pm)}{c_0^2} \\
&&= \mathcal{W }(0,0,0) \pm \frac{1}{2}\Omega \partial_\omega \mathcal{W } \pm \frac{1}{2}\mathbf{q}\cdot \partial_\mathbf{k} \mathcal{W } \pm \frac{4\pi i\sigma }{c_0^2}+ \cdots
\end{eqnarray*}
and set $\Delta W =  \mathcal{W}_+ - \mathcal{W}_-$ . With this notation the noise correlation in Eq.~(\ref{fd}) takes the form $J_{jl}(\omega,q,\Omega)     = Q(\omega) $  $ \sigma_{jl}(\omega,\mathbf{q})\delta_\Omega$.  To be able to identify energy and energy current we consider  first an arbitrary  current correlation function $J_{jl,\omega} (\mathbf{q}, \Omega)$ and homogeneous dissipation $\sigma_{jl}(\omega)$ and take the limit $\Omega \rightarrow 0$ towards a stationary process afterwards.
Similarly, the limit $\mathbf{q}\rightarrow 0$ anticipates the noise to be uniformly present throughout the medium. In that case the radiation intensity, averaged over the random source, is given by
\begin{eqnarray*}
  \phi_{ik}(\Omega,\mathbf{k},\mathbf{q})   = G^s_{ij}(\omega_+,\mathbf{k}_+) J_{jl,\omega}(\mathbf{q}, \Omega) G^{s\dag}_{lk}(\omega_-,\mathbf{k}_-)
\end{eqnarray*}
and we can write
\begin{eqnarray*}
 && \frac{c_0^2}{8\pi}\Delta \mathcal{W}_{ki} (\Omega, \mathbf{k}, \mathbf{q}) \phi_{ik,\mathbf{ k}}(\Omega,\mathbf{q})  =  \\
 &&  = \frac{2\pi}{c_0^2} G^{s\dag}_{lk}(\omega_-,\mathbf{k}_- ) \Delta \mathcal{W}_{ki}(\Omega, \mathbf{k}, \mathbf{q}) G^s_{ij}(\omega_+,\mathbf{k}_+) J_{jl,\omega} (\mathbf{q}, \Omega) \\
  && =\frac{2\pi}{c_0^2}\left[ G^{s\dag}_{lj}(\omega_-,\mathbf{k}_- ) - G^s_{lj}(\omega_+,\mathbf{k}_+) \right] J_{jl,\omega} (\mathbf{q}, \Omega)
\end{eqnarray*}
Upon integrating over all wave vectors $\mathbf{k}$ and in the limit $\Omega,\mathbf{q} \rightarrow 0$;  this can be seen to be equivalent  to a continuity equation $-i\Omega \mathcal{E} -i\mathbf{q}\cdot \mathbf{S} + A = J $.  The extra factor $c_0^2/8\pi$ guarantees that $\mathcal{E} (\Omega, \mathbf{q}) (d\omega/2\pi) (d\Omega/2\pi) d^3\mathbf{q}/(2\pi)^3$ coincides with the usual energy density with the correct dimension, $\partial_\omega \mathcal{W}$ having the dimension $c_0^{-2}$. This identifies (average) energy density and energy current density in Fourier space as
\begin{eqnarray}
  \mathcal{E}(\Omega,\mathbf{q}) &=& \frac{c_0^2}{8\pi}\mathrm{Tr} \sum_\mathbf{k} \partial_\omega \mathcal{W }(\mathbf{k},\mathbf{q}) \cdot \phi_\mathbf{k} (\Omega, \mathbf{q}) \label{ES1} \\
  \mathbf{S}(\Omega,\mathbf{q}) &=& \frac{c_0^2}{8\pi} \mathrm{Tr} \sum_\mathbf{k} \partial_\mathbf{k }\mathcal{W }(\mathbf{k},\mathbf{q}) \cdot \phi_\mathbf{k}(\Omega, \mathbf{q}) \label{ES2}
\end{eqnarray}
  The absorption rate is identified as
\begin{eqnarray}
a (\mathbf{q},\Omega) =   \mathrm{Tr} \sum_\mathbf{k} \sigma  \cdot \phi_\mathbf{k} (\Omega,  \mathbf{q})
\end{eqnarray}
which we recognize as the Ohmic dissipation $A d\omega\sim \sigma E^2$. Finally ,the emission rate of the source is
 \begin{eqnarray}
\nonumber e (\mathbf{q},\Omega)=  \frac{2\pi}{ ic_0^2}\mathrm{Tr} \sum_\mathbf{k}\left[ \mathbf{G}^{s\dag} (\omega_-,\mathbf{k}_- ) - \mathbf{G}^s(\omega_+,\mathbf{k}_+) \right] \cdot \mathbf{J}_{\omega} (\mathbf{q}, \Omega)\\
\end{eqnarray}
The equations~(\ref{ES1}) and (\ref{ES2}) assume the medium to be homogeneous and conservative, and allow dispersion in both $\omega$ and $\mathbf{k}$, although the latter has not been treated here explicitly. For illustration, for isotropic, dispersive and conservative media we recover the well-known expression

 \begin{equation}
 \mathcal{ E}(\omega) = \frac{\mathbf{E}(\omega) \cdot \overline{\mathbf{E}}(\omega)}{8\pi }\frac{d}{d\omega}\omega\epsilon(\omega)  - \frac{\mathbf{B} (\omega) \cdot \overline{\mathbf{B}}(\omega)}{8\pi } \omega^2\frac{d}{d\omega}\frac{1}{\omega\mu(\omega)}
\end{equation}
for the spectral energy density, where we used $\omega \mathbf{B} = c_0(\epsilon\cdot \mathbf{p})\cdot \mathbf{E}$. As is well known, the spectral energy density is affected by frequency dispersion, unlike the spectral energy current density $\mathbf{S}$ in Eq.~(\ref{ES2}). Even in the presence of bi-anisotropy,  $\mathbf{S}$ coincides with the  Poynting vector $\mathbf{E}(\omega) \times \mathbf{\bar{H}}(\omega)/8\pi + c.c. $. In homogeneous, conservative and local dielectric media the Poynting vector is parallel to the group velocity \cite{LLgroup} which excludes the well-known ambiguity to add any curl of a vector field to the Poynting vector to describe the energy current \cite{jackson}. Only when we would allow spatial dispersion explicitly in the dielectric tensor $\varepsilon(\mathbf{k})$, the Poynting vector fails to describe the energy current, and an additional  contribution  proportional to $(\partial_\mathbf{k} \varepsilon_{nm}) E_n\bar{E}_m $ from the material enters the energy current density \cite{maddox,Alex}.

If we now restrict to stationary and homogeneous noise fluctuations, we impose $\mathbf{J}(\mathbf{q},\Omega) = \mathbf{J} \delta_\mathbf{q}\delta_\Omega$. Upon applying Eq.~(\ref{fd}), we see that $a=e$, provided that

\begin{eqnarray} \label{DeltaG}
 \phi_\mathbf{k}(\omega, \mathbf{q}, \Omega ) =  \frac{ 4\pi \omega Q(\omega) }{ic_0^2}\left[ \mathbf{G}^\dag (\omega ,\mathbf{k}  ) - \mathbf{G}(\omega ,\mathbf{k} ) \right] \delta_\mathbf{q} \delta_\Omega
\end{eqnarray}
which is just the equivalent of Eq.~(\ref{GG}) in a homogeneous medium. Hence, the detailed balance of source and absorption guarantees the radiation energy to be constant in time.

 In the following we will treat a well-known feature, equipartition of energy, on the same basis as a zero energy current.
In a bi-anisotropic medium, this relation is non-trivial since  $  \phi_{ij, \mathbf{k}} \neq  \phi_{ij, -\mathbf{k} }$.

\subsubsection{Equipartition}
For an energy distribution in phase space given by Eq.~(\ref{DeltaG}),  Eq.~(\ref{ES1}) yields the stationary and homogenous spectral energy density,

\begin{equation}\label{E4}
 \mathcal{E}(\omega,\mathbf{r},t) = \frac{Q(\omega)}{2i}\mathrm{Tr} \sum_\mathbf{k }  \partial_\omega \mathcal{ W} \cdot \left[ \mathbf{G}^{s\dag}(\omega,\mathbf{k}) - \mathbf{G}^s(\omega,\mathbf{k})\right]
\end{equation}
We will suppose that the dissipation is small, so that $4\pi i\sigma/c_0^2\rightarrow i\epsilon$ and $\mathbf{G}^s(\omega,\mathbf{k})= (\mathcal{W}(\omega,\mathbf{k}) + i\epsilon)^{-1}$.
Since  $\mathcal{W}$ is hermitian, is has 3 real-valued eigenvalues written as $w_{\mathbf{g}}(\omega,\mathbf{k})$. \emph{Provided } we continue the logarithm analytically into the complex plane with a branch cut located at the positive real axis, we find
\begin{eqnarray*}
  \mathrm{Tr} \,\partial_\omega \mathcal{W}\cdot  \left(\mathcal{W} (\omega,\mathbf{k}) \pm i\epsilon\right)^{-1}
  = \sum_\mathbf{g} \, \partial_\omega  \log \left(w_\mathbf{g}(\omega,\mathbf{k})   \pm i\epsilon \right)
\end{eqnarray*}
 Hence
\begin{eqnarray}
\nonumber  \mathcal{E}_0(\omega) =  \frac{Q(\omega)}{2  i}\partial_\omega\sum_\mathbf{kg }   \left[ \log \left( w_{\mathbf{g}}(\omega,\mathbf{k})-i\epsilon\right) \right. \\
\nonumber \left. -\log \left(w_{\mathbf{g}}(\omega,\mathbf{k}) +i\epsilon \right)
\right]
\end{eqnarray}
The difference between the logarithms is  non-zero and equal to $2\pi i$ only for wave vectors such that $w_{\mathbf{g}}(\omega,\mathbf{k})>0 $. If we denote by $\omega_\mathbf{g}(\mathbf{k})$ the solution of $w_{\mathbf{g}}(\omega,\mathbf{k})=0 $ and define all $\mathbf{k}$-vectors satisfying this equation for fixed $\omega$ as the constant-frequency surface $\mathbf{S}_{\omega, \mathbf{g}} $, we can write the energy density as
\begin{eqnarray}\label{EP}
   \nonumber  \mathcal{E}_0(\omega)&=&  {\pi Q(\omega) } \partial_\omega \sum_{\mathbf{k g}} \{w_{\mathbf{g}}(\omega,\mathbf{k})>0\} \\ &=& { \pi Q(\omega)} \frac{1}{(2\pi)^3}\sum_\mathbf{g} \int \frac{d^2S_{\omega, \mathbf{g}}}{|d\omega_{\mathbf{gk}}/d\mathbf{k}|}
\end{eqnarray}
The first equality illustrates equipartition: the energy  $  \mathcal{E}(\omega)\times d\omega/2\pi $ per unit volume is just proportional to the total number of states available per unit volume in the frequency interval $d\omega $ in phase space, each having the energy $Q/2$.  The second equality assumes that $w_{\mathbf{g}}(\omega,\mathbf{k})>0$ implies $\omega> \omega_\mathbf{g}(\mathbf{k})$ (positive group velocity)  and leads to
a well-known text-book expression for the density of states per unit volume with the group velocity $d\omega_{\mathbf{gk}}/d\mathbf{k}$ as the vector normal to $\mathbf{S}_{\omega, \mathbf{g}}(\omega)$. If an eigenvalue $w_\mathbf{g}(\omega,\mathbf{k})  $ independent on $k$ is encountered, typically associated with a longitudinal excitation, group velocity and constant-frequency surface are ill-defined , yet the first line of Eq.~(\ref{EP}) shows that this eigenvalue does not contribute to $ \mathcal{E}(\omega)$.

\subsubsection{Vanishing of Current}
The energy current density can be treated in exactly the same way,

\begin{eqnarray}
\nonumber \langle\mathbf{S}_0(\omega)\rangle =  \frac{Q(\omega)}{ 2i}\sum_\mathbf{kg} \partial_\mathbf{k }  \left[ \log  \left({ w }_\mathbf{g}(\omega,\mathbf{k}) -i\epsilon  \right) \right. \\
\nonumber \left. -\log  \left({ w}_\mathbf{g} (\omega,\mathbf{k}) +i\epsilon  \right)
\right] \end{eqnarray}
Applying the gradient theorem in vector calculus,
\begin{eqnarray}\label{shit3}
  &&\nonumber\langle\mathbf{S}_0(\omega) \rangle = \frac{Q(\omega)}{2i}\frac{1}{(2\pi)^3}\lim_{k\rightarrow\infty}
k^2\int d^2\mathbf{\hat{k}} \, \mathbf{\hat{k}} \sum_\mathbf{g} \\ &&\nonumber \left[
\log  \left(\frac{1}{k^2} { w }_\mathbf{g}(\omega,\mathbf{k}) -i\epsilon  \right) -\log  \left(\frac{1}{k^2} { w}_\mathbf{g }(\omega,\mathbf{k}) +i\epsilon  \right)
\right]\\
\end{eqnarray}
where we divided by $k^2$ for convenience, $\log k$ canceling in the expression;
$ { w }_\mathbf{g}(\omega,\mathbf{k}) $ are the eigenvalues of the hermitian matrix
 \begin{equation}
     \hat{ {W}}\equiv \frac{1}{k^2} {W}= \frac{1}{\omega}\varepsilon_\mathbf{\hat{k}} \cdot\mu^{-1}\varepsilon_\mathbf{\hat{k}} + \frac{\omega }{k^2c_0^2} \mathcal{E} + \frac{1}{kc_0}\left(\varepsilon_\mathbf{\hat{k}} \cdot \chi^* - \chi\cdot \varepsilon_\mathbf{\hat{k}}\right)
\end{equation}
For large $k$, we investigate how they are affected by the last bi-anisotropic term of $ \hat{ {W}}$, which is  is odd in $\mathbf{k}$, and without which the $\mathbf{\hat{k}}$-integral for $\mathbf{S}_0$ would trivially vanish. For a positive eigenvalue the difference of the two logarithms equals $2\pi i$, for a negative eigenvalue the two logarithms cancel. The first term has 3 real-valued eigenvalues $w_{ {\mathbf{g}}}(\omega,\mathbf{k})$, 2 that are finite and negative under normal conditions for $\mu$(excluding meta-materials where $\mu $ can be negative, which would  necessarily be accompanied with dispersion and absorption) with transverse eigenvectors $\mathbf{g}_T^{(1,2)}$ and one eigenvalue 0 with longitudinal eigenvector $\mathbf{g}_L = \hat{\mathbf{k}}$. As $k\rightarrow\infty$, the bi-anisotropic will just give a vanishing perturbation to the first two non-zero eigenvalues and cannot change their sign as $\mathbf{\hat{k}}$ varies. Hence, they give no contribution to $\mathbf{S}_0$.
The zero eigenvalue is perturbed linearly by the dielectric term $\omega \varepsilon/k^2c_0^2$, and the bi-anisotropic term comes in only into second order perturbation theory,
 \begin{eqnarray*}
      w_{{\mathbf{g}_L}}(\omega,\mathbf{k}) = \frac{1}{k^2}\left[\frac{\omega}{c_0^2}\langle \mathbf{\hat{k}}|  \mathcal{E} (\omega) |\mathbf{\hat{k}}\rangle+ \frac{4}{c_0^2}\sum_{i=1,2} \frac{|\langle \mathbf{g}_T^{(i)}| \varepsilon_\mathbf{k} \cdot \chi^* | \mathbf{\hat{k}} \rangle|^2}{w_{{\mathbf{g}}_T^{(i)}}(\omega,\mathbf{k})} \right]
 \end{eqnarray*}
It vanishes as $1/k^2$ and, more importantly, is invariant under $\mathbf{\hat{k}} \rightarrow -\mathbf{\hat{k}}$. Therefore, again, the vector-surface integral in Eq.~(\ref{shit3}) vanishes. We conclude that under very general conditions,
\begin{equation}\label{shit2}
  \langle\mathbf{S}_0(\omega) \rangle = 0
\end{equation}

\subsection{Momentum Density }

The electromagnetic  fields $\mathbf{E}$ and $\mathbf{B}$ as well as the Lorentz force couple to the matter via the charge density $\rho$ and the current density $\mathbf{J}$. To obtain a macroscopic picture one expands both in gradients. If no net local charges and currents are present, the induced charge density is written as $\rho = - \nabla\cdot \mathbf{P}$ and the induced current density as $\mathbf{J}= \partial_t \mathbf{P} + c_0\nabla\times \mathbf{M}$. The Lorentz force density is given by
${d \rho\mathbf{v}}/{dt}= \rho \mathbf{E} + \mathbf{J} \times \mathbf{B }/c_0$,
with $\mathbf{v}$ the local velocity. After some vector calculus with  no further assumptions, this equation can be transformed into one that transfers momentum  from  the  electromagnetic field to the particle,

\begin{eqnarray}\label{mm1}
    \nonumber \frac{d\rho\mathbf{v}}{dt} &=& \mathbf{f}_A + \mathbf{f}_C \\
    &-&\nabla\cdot \left\{\mathbf{PE} -\mathbf{BM} - \frac{{1}}{2} (\mathbf{P}\cdot \mathbf{E} - \mathbf{B}\cdot \mathbf{M})\right\}
\end{eqnarray}
The  force density $\mathbf{f}_C$ exerted by the electromagnetic field on the matter is given by
\begin{equation}\label{fM}
   \mathbf{f}_C = \frac{1}{2}\nabla(\mathbf{P}\cdot \mathbf{E}) + \frac{1}{2}\nabla(\mathbf{M}\cdot \mathbf{B}) + P_m \nabla E_m + M_m \nabla B_m
\end{equation}
and involves spatial derivatives. The force density $\mathbf{f}_A$  involves a time derivative,
\begin{equation}\label{fE}
   \mathbf{f}_A = \frac{1}{c_0}\partial_t \left( \mathbf{P}\times \mathbf{B}\right)
\end{equation}
and is usually referred to as the ``Abraham force density".  A second equation can be obtained directly from Maxwell equations, that describes the transfer of momentum from matter to radiation,
\begin{eqnarray}\label{mm2}
   \nonumber \frac{1}{4\pi c_0}\partial_t \left( \mathbf{E}\times \mathbf{B}\right)  &=&  -\mathbf{f}_P -\mathbf{f}_C  \\
     \nonumber  +\frac{1}{4\pi }\nabla &\cdot& \left\{\mathbf{DE} +\mathbf{BH} - \frac{{1}}{2} \left(\mathbf{D}\cdot \mathbf{E} + \mathbf{B}\cdot \mathbf{H}\right)\right\}\\
\end{eqnarray}
where, as usual, $\mathbf{D} = \mathbf{E} + 4 \pi \mathbf{P} $ and $\mathbf{H }= \mathbf{B} - 4\pi \mathbf{M}$. Written in this form, Newton's third law is manifestly obeyed and the  electromagnetic momentum density is identified as $\mathbf{K}= (4\pi c_0)^{-1}\mathbf{E} \times \mathbf{B} $. Upon adding Eq.~(\ref{mm1}) and (\ref{mm2})  we establish conservation of ``total" momentum,

\begin{eqnarray}\label{mm3}
 \partial_t \mathbf{K} & + & \frac{ d\rho\mathbf{v}}{dt} = \nabla\cdot \mathbf{T}
\end{eqnarray}
where
 \begin{eqnarray}\label{TN}
\mathbf{ T}=\frac{1}{4\pi } \cdot  \left\{\mathbf{E {E}} +\mathbf{B {B}} - \frac{\mathbf{1}}{2} \left(\mathbf{E}\cdot \mathbf{ {E}} + \mathbf{B}\cdot \mathbf{ {B}}\right)\right\}
 \end{eqnarray}
is recognized as the ``vacuum" momentum stress tensor, whose expression is symmetric.
Equation~(\ref{mm3}) agrees with Ref.~\cite{nelson}, and is equal to the microscopic momentum balance in vacuum which does not feature \emph{explicitly} any constitutive parameter. It will therefore not be modified by either spatial or dynamic dispersion, and also remains valid if the constitutive equations~(\ref{CE}) contain an explicit time-dependence. This is not true for e.g. the Minkowski momentum density $\mathbf{D }\times \mathbf{B}/4\pi c_0 $ which is affected by dispersion when expressed in the fields $\mathbf{E}$ and $\mathbf{B}$ \cite{Bliokh2}.

The simplest case to consider is a homogeneous medium filled with an on average homogeneous random source, for which the gradient on the righthand side must disappear. We will consider the more realistic case of a finite medium in the last section, and show that the momentum leak through a surrounding surface vanishes. Here we have,

\begin{equation}\label{mm4}
   \frac{d  \rho\mathbf{v}}{dt}+ \frac{d\langle \mathbf{K} \rangle}{dt}=0
\end{equation}
It is convenient to symmetrize when going over to Fourier space,

\begin{equation}\label{Ksymm}
    \mathbf{K}(\omega) = \frac{1}{8\pi c_0} \left( \mathbf{E} \times \mathbf{\overline{B}} - \mathbf{B} \times \mathbf{\overline{E}} \right).
\end{equation}
and to use that $\langle\mathbf{E} \times \mathbf{\overline{H}}\rangle =0 $. To simplify the analysis we will assume $\mu$ to be a real-valued scalar. If we insert $\mathbf{B}= \mu\mathbf{H} + \mu \chi^*\cdot \mathbf{E}$ it follows
\begin{eqnarray}\label{Ksymm2}
\nonumber \langle K_i(\omega)\rangle &= &-\frac{\mu}{4\pi c_0}\mathrm{ Re} \, \sum_\mathbf{k} \mathrm{Tr} \left (\epsilon_i\cdot \chi^*\cdot \phi_\mathbf{k} \right)  \\
&=& - \frac{\mu}{4\pi c_0}\mathrm{ Re} \, \mathrm{Tr} \left (\epsilon_i\cdot \chi^*\cdot \mathbf{A}  \right)
\end{eqnarray}
with $(\epsilon_i)_{jk}= \epsilon_{jki}$, $\phi_\mathbf{k}(\omega)$ given by Eq.~(\ref{DeltaG}), and whose $\mathbf{k}$-integral is set equal to the hermitian matrix $\mathbf{A}(\omega)$. Since $\chi$ is usually very small we can neglect any bi-anisotropic behavior in $\mathbf{A}(\omega)$.

For the momentum density to be non-zero we need that $\chi^*\cdot\mathbf{ A}$ be an anti-symmetric matrix. We will discuss two possibilities. The first concerns an optical material  that exhibits both rotatory power and Faraday rotation, both well-known sources of circular dichroism (CD).  In that case $\chi = ig(\omega)c/\omega  $  with $g$ a real-valued pseudo-scalar,  indicating the amount of rotation per meter caused by microscopic chirality, and $\varepsilon_{ij} = \varepsilon (\omega)\delta_{ij} + i (c/\omega) V(\omega) \epsilon_{ijk}B_{0,k}$, with the Faraday effect quantized by Verdet constant which measures the amount of rotation per meter per Tesla ($c=c_0/\sqrt{\varepsilon\mu})$. It follows that

\begin{eqnarray*}
  K_i(\omega) = \frac{\omega Q(\omega) g(\omega)\mu(\omega)}{c_0^2} \mathrm{Re} \, \mathrm{Tr} \, \epsilon_i\cdot \sum_\mathbf{k}(\delta \mathbf{G}^\dag - \delta \mathbf{G})
\end{eqnarray*}
with $\mathbf{G}$ the Green's function of the homogeneous Helmholtz equation. Linearizing in the external magnetic field yields,

\begin{eqnarray*}
    &&\sum_\mathbf{k}\mathbf{ }\delta \mathbf{G}(\mathbf{k},\omega,\mathbf{B}_0) = -\frac{\mu^2  \omega V}{c} \times \\
    &&  \sum_\mathbf{k} \frac{1}{\omega^2/c^2 - k^2 +\mathbf{kk} +i\epsilon} \cdot (i\epsilon\cdot \mathbf{B}_0) \cdot  \frac{1}{\omega^2/c^2 - k^2 +\mathbf{kk} +i \epsilon} \\
    && = \left(i\frac{\mu^2 V }{8\pi} + \Lambda \right)(i\epsilon\cdot \mathbf{B}_0)
\end{eqnarray*}
Here, $\Lambda$ is a real-valued positive cut-off associated with the diverging longitudinal field but cancels upon subtracting the hermitian conjugate. Thus the electromagnetic momentum density is $\mathbf{K}(\omega)\,  d \omega/2\pi $ with
\begin{eqnarray}\label{KgB}
  \mathbf{K}(\omega) = \frac{Q(\omega)}{2\pi}\frac{\mu^2 (\omega)    g(\omega)  V(\omega) }{ c_0^2} \mathbf{B}_0
\end{eqnarray}
The simultaneous action of two well-known mechanisms for CD leads to a finite electromagnetic momentum density  accumulated in the material.  Any small variation of the external magnetic field leads to a force $-d\mathbf{K}/dt$ per unit volume exerted on the material, as expressed by Eq.~(\ref{mm4}). Optical forces induced by time-dependent electromagnetic fields appear in Eq.~(\ref{mm1}). However, an optical "Abraham force"  proportional to $gV d\mathbf{B}/dt $ does not show up in Eq.~(\ref{mm1}) and is here generated by the presence of the radiation noise. If we would apply  the above formula for $\mathbf{K}$ to the quantum vacuum ($Q = \hbar \omega$), using that at high frequencies $V(\omega)\sim 1/\omega^2$ if the Verdet constant $V$ is associated with the microscopic Zeeman shift of atomic levels, and similarly $g(\omega)\sim 1/ \omega^2$ if $g$ is associated with a simultaneous magnetic and electric dipole transition in a chiral molecule \cite{Craig8-5}, the frequency integral does not seem to suffer from an ultraviolet catastrophe. A realistic calculation should however involve QED. A quantum-mechanical treatment of a chiral molecule with Zeeman effect interacting with the quantum vacuum predicts indeed a finite effect with velocities of order $10^{-9}$ m/s \cite{donaire}. A more phenomenological treatment based on magneto-chiral anisotropy in the Einstein coefficients even yields velocities of the order of $1 \, \mu$m/s \cite{georges}.  For classical thermal noise with bandwidth $kT/h$ ($\approx 10^{13}$ Hz at room temperature) this effect is entirely negligible:  with $VB= 100$ rad/$m$, $g= 3\cdot 10^{-2} $ degrees per wavelength, $\rho = 4 \, g/cm^3$, Eqn.~(\ref{KgB}) yields at room temperature ($Q \approx 0.03$ eV) a velocity $v= 10^{-25} \, m/s$.   Values of $200$ degrees per wavelength have been reported for meta-materials in the THz region \cite{meta1}, even $450$ degrees per wavelength around $100$ THz \cite{meta2}. When we would expose them to isotropic radiation, for which  $Q= Fc_0^2/\omega^2 $ can be much larger ($10^7$ eV for light fluxes of $100$ kW/$m^2$ over a bandwidth of $100$ kHz in the visible, the predicted radiation momentum is hardly much larger, essentially due to the much smaller bandwidth. We conclude that the observation of radiation momentum induced by classical noise remains a thought experiment.

The second application is obtained for $ \mathbf{\chi}^*=  \epsilon \cdot \mathbf{w} $, with $\mathbf{w}$ some real-valued dimensionless vector, odd in both time-reversal and parity. From Eq.~(\ref{Gr}), we see that the origin of the Green function in phase space shifts from $\mathbf{k}=0 $ to  $\mathbf{k}= -\mathbf{w }$. Since this is also true for the object $\mathcal{W}$ in Eqs.~(\ref{ES2}), the Poynting vector can immediately be seen to vanish . If we adopt scalar $\varepsilon $ and $\mu$, we have

\begin{eqnarray*}
  K_i(\omega) = -\frac{\omega\mu^2 Q }{ i c_0^3} \mathrm{Tr} \, \epsilon_i\cdot \sum_\mathbf{k}(\mathbf{G}^* - \mathbf{G})
\end{eqnarray*}
With
\begin{eqnarray*}
  \sum_\mathbf{k} \mathbf{G}  =  \sum_\mathbf{k} \frac{1}{\omega^2/c^2 - k^2 +\mathbf{kk} } = -\frac{i}{3\pi } \frac{\omega}{c} + \Lambda
\end{eqnarray*}
Since $\Lambda$ again cancels in $\sum_\mathbf{k} (\mathbf{G }- \mathbf{G}^\dag) $, we find
\begin{eqnarray*}
  \mathbf{K}(\omega) = -\frac{2Q(\omega)}{3\pi}\frac{\mu^2(\omega)     \omega^2 }{ c c_0^3} \mathbf{w}
\end{eqnarray*}
This is basically a result first due to Feigel \cite{feigel}, who applied it to the quantum vacuum with  magneto-electric anisotropy for which $\mathbf{w}= g_{ME} \mathbf{E}_0 \times \mathbf{B}_0$. In this case the momentum is directed along the vector $ \mathbf{E}_0 \times \mathbf{B}_0$ and the force proportional to $d/dt (\mathbf{E}_0 \times \mathbf{B}_0) $. This can be seen as a correction to the Abraham force, obtained from Eq.~(\ref{fE}).

\section{Bounded Media}

In the following we consider a bounded, bi-anisotropic medium, emerged in an infinite sea of homogeneous fluctuations described by Eq.~(\ref{fd}) with homogeneous  $Q$. As a result, Eq.~(\ref{DeltaG}) applies in- and outside the medium. This situation can be easily seen to be equivalent to a  homogeneous collection of uncorrelated random sources in the far field of the medium.

\subsection{No net Poynting vector}
Despite being intuitively plausible, \emph{the average Poynting vector does not rigorously vanish {everywhere} in a bounded, heterogeneous medium} in the presence of an isotropic noise field, that is, if Eq.~(\ref{DeltaG}) is satisfied.  What is well-know is that \cite{jackson}
\begin{eqnarray*}
  \nabla\cdot \mathbf{S} &=& -\mathbf{J}\cdot \mathbf{E}
\end{eqnarray*}
with $\mathbf{S}= c_0\mathbf{E} \times \mathbf{H}/4\pi $ the local Poynting vector. When applied to the noise,
\begin{eqnarray}\label{Sisnul}
\nonumber  \nabla\cdot \langle\mathbf{S}\rangle &=& -\sigma_{ij}(\mathbf{r})\langle {\overline{E}}_i(\mathbf{r})E_j(\mathbf{r})\rangle + \mathrm{Re }\, \frac{4\pi i \omega }{c_0^2} G_{ij}(\mathbf{r},\mathbf{r}) J_{ji}(\mathbf{r}) \\ &=&  0
\end{eqnarray}
 if the fluctuation-dissipation theorem~(\ref{fd}) is satisfied , and in which case Eq.~(\ref{GG}) applies. The vanishing divergence of $\mathbf{S}$ alone does not make $\langle\mathbf{S}  \rangle$ vanish itself. The  Helmholtz theorem in vector calculus states that under very broad conditions, $\mathbf{S}$ must be equal to the curl of a vector field $\mathbf{V}(\mathbf{r})$. In the language of Ref.~\cite{Bliokh2} this implies that the ``canonical" momentum vanishes.   What is assumed in Ref.~\cite{Bliokh2} on the basis of ``localized fields vanishing at infinity", is proven here more explicitly,
\begin{eqnarray}\label{Sint}
 \int d^3 \mathbf{x} \langle \mathbf{S}(\mathbf{x}) \rangle =0
 \end{eqnarray}
The vanishing of the \emph{integrated} Poynting vector was previously demonstrated numerically for a set of $N$ electric dipoles with Zeeman effect \cite{pinheiro}, and where PT-symmetry considerations would allow a net current proportional to $g \mathbf{B}$  to  occur, with $g$ a pseudo scalar associated with a chiral configuration of the $N$ dipoles. Being the curl of a vector field, a surface term may still survive in Eq.~(\ref{Sint}). In the next section we will show that the angular momentum $\int d^3 \mathbf{x} \,  \mathbf{x}\times \langle \mathbf{S} \rangle $ does not generally vanish.

The Poynting vector, in complex frequency notation, is given by $\mathbf{S}= c_0 [ \overline{\mathbf{E}} \times \mathbf{H} - \overline{\mathbf{H}} \times \mathbf{E}]/8\pi$. Since $\partial_\mathbf{p}\mathcal{W}_{ij}(\mathbf{r},\mathbf{p}) = -\epsilon_{iln}(\mu^{-1}\cdot \mathbf{L})_{lj}+ (\mu^{-1}\cdot \mathbf{L})^\dag_{il}\epsilon_{ljn}$, with $\mathbf{L}$ defined Eq.~(\ref{Gr}), and $\mathcal{W} = \omega^2 \mathcal{E}/c_0^2 + \mathbf{L}^\dag\cdot \mu(\mathbf{r})^{-1} \cdot \mathbf{L}$ the hermitian operator defined in section 2, but here spatially inhomogeneous, it follows that
 \begin{eqnarray*}
    {S}_n(\omega,\mathbf{r}) =  -\frac{c^2_0}{8\pi \omega} \overline{E}_i(\mathbf{r})  \frac{\partial \mathcal{W}_{ij}}{\partial p_n} {E}_i(\mathbf{r}) + c.c.
 \end{eqnarray*}
We can check that
\begin{eqnarray}
 \nonumber &&\int d^3 \mathbf{x } \langle \mathbf{S}(\mathbf{x}) \rangle  = \int d^3 \mathbf{x}  \langle \mathbf{x}| {\mathbf{S}}(\omega,\mathbf{ p},\mathbf{r}) |\mathbf{x}\rangle = \sum_\mathbf{k} \langle \mathbf{k}|\langle \mathbf{S} \rangle  |\mathbf{k}\rangle \\
  \nonumber &&= -\frac{Q}{2 i} \sum_\mathbf{k} \langle \mathbf{k}| \mathrm{Tr} \left(\frac{\partial}{\partial\mathbf{p}} \mathcal{W}((\mathbf{r},\mathbf{p})\right)  \cdot [\mathbf{G}(\mathbf{r},\mathbf{p}) -\mathbf{G}^\dag(\mathbf{r},\mathbf{p}) ] | \mathbf{k}\rangle \\
 \nonumber &&= -\frac{Q}{2 i}  \mathrm{TR}  \left(\frac{\partial}{\partial\mathbf{p}} \mathcal{W}((\mathbf{r},\mathbf{p})\right)\cdot [\mathbf{G}(\mathbf{r},\mathbf{p}) -\mathbf{G}^\dag(\mathbf{r},\mathbf{p}) ]
\end{eqnarray}
where $\mathrm{Tr}$ stands for trace over polarization indices, and $\mathrm{TR}$ the full trace in Hilbert space. We have $\mathbf{G} = (\mathcal{W} + i \epsilon)^{-1}$. The operator
 $ {\partial}\mathcal{W}/{\partial\mathbf{p}} $ does not necessarily commute with $\mathcal{W}$ itself. The cyclic property of the full trace allows to write
\begin{eqnarray}
\nonumber \int d^3 \mathbf{x} \langle \mathbf{S}(\mathbf{x}) \rangle =
 -\frac{Q}{2 i}  \mathrm{TR} \frac{\partial}{\partial\mathbf{p}} \left[ \log(\mathcal{W} + i \epsilon) -  \log(\mathcal{W} - i \epsilon) \right]
 \end{eqnarray}
Finally we have in general

 \begin{eqnarray}
 \nonumber \langle \mathbf{k}|  \frac{\partial}{\partial\mathbf{p}}  {A}(\mathbf{r},\mathbf{p}) | \mathbf{k}\rangle=
-\frac{1}{i}  \langle \mathbf{k}| [A(\mathbf{r},\mathbf{p}), \mathbf{r}]   | \mathbf{k}\rangle = \frac{d}{d\mathbf{k}}  \langle \mathbf{k}| A(\mathbf{r},\mathbf{p})  | \mathbf{k}\rangle
\end{eqnarray}
Applying the gradient theorem leads to an expression similar to the one found earlier for a homogeneous medium,
\begin{eqnarray}
 \nonumber &&\int d^3 \mathbf{x } \langle \mathbf{S}(\mathbf{x}) \rangle = -\frac{Q}{2i} \frac{1}{(2\pi)^{3}}  \\
 \nonumber && \times \lim_{k\rightarrow \infty} k^2 \int d^2\hat{\mathbf{k}}\hat{\mathbf{ k}}
 \langle \mathbf{k}|   \left[ \log(\mathcal{W} + i \epsilon) -  \log(\mathcal{W} - i \epsilon) \right] | \mathbf{k}\rangle
\end{eqnarray}
The large wave number limit makes us enter into the regime of geometrical optics, yet here at fixed frequency. If the wavelength is smaller than the typical spatial variations in the material parameters, the notion of a locally homogeneous medium applies, in which case the commutation relation between $\mathbf{r}$ and $\mathbf{p}$ can be neglected. As a result,
 \begin{eqnarray}
 \nonumber \langle \mathbf{k}|    {A}(\mathbf{r},\mathbf{p}) | \mathbf{k}\rangle \rightarrow
  \int d^3 \mathbf{x}   A(\mathbf{x},\mathbf{k})
\end{eqnarray}
 as $k\rightarrow \infty$, to find,
  \begin{eqnarray}
 \nonumber &&\int d^3 \mathbf{x } \langle \mathbf{S}(\mathbf{x}) \rangle = -\frac{Q}{2 i} \frac{1}{(2\pi)^{3}}  \times \int d^3 \mathbf{x } \\
 \nonumber && \lim_{k\rightarrow \infty} k^2 \int d^2\hat{\mathbf{k}}\, \hat{\mathbf{ k}}
    \left[ \log(\mathcal{W}(\mathbf{x},\mathbf{k}) + i \epsilon) -  \log(\mathcal{W}(\mathbf{x},\mathbf{k})  - i \epsilon) \right] |
\end{eqnarray}
The operator $\mathcal{W}(\mathbf{r},\mathbf{p})$ is here reduced to a real-valued inhomogeneous number. For every $\mathbf{x}$, the surface integral over $\hat{\mathbf{k}}$  vanishes everywhere just as it vanished for a translationally invariant medium.

The "no-net current" theorem has no direct consequences for local energy balance and is consistent with $\mathbf{S}$ being a curl. However, it implies immediately that if $\mathbf{B}=\mathbf{H}$ everywhere in space, the total radiation momentum $\int d^3 \mathbf{x} \, \langle\mathbf{K}(\mathbf{x} \rangle) $ induced by the random sources vanishes rigorously, and that the average angular momentum $\int d^3 \mathbf{x} \, \mathbf{x}\times \, \langle\mathbf{K}(\mathbf{x} \rangle) $ does not depend on the chosen origin of the integral.

\subsection{No net momentum leaks}
In a finite medium, total momentum is conserved unless momentum leaks infinity. This is expressed by,
\begin{eqnarray}\label{mm3a}
   \nonumber   \partial_t \int d^3\mathbf{r} \, \mathbf{K}(\mathbf{r})  +  m\frac{d\mathbf{v}}{dt} =
 \nonumber  \lim_{r\rightarrow\infty}{r^2}\int d^2 \hat{\mathbf{r}}\, \hat{\mathbf{ r}} \cdot \mathbf{T}
\end{eqnarray}
 with the momentum-stress tensor $\mathbf{T}$ defined in Eq.~(\ref{TN}), and where we assume that the spherical boundary is located in the far field of the medium. The ''Im $G$ "-theorem~(\ref{GG}) applies also at this boundary. In the far field we can write,

 \begin{eqnarray}\label{Gff}
\nonumber && \mathbf{G}(\mathbf{r},\mathbf{r},\omega)=   \mathbf{G}_{0}(0,\omega) \\
\nonumber &+&   \int d^3\mathbf{x} \int d^3\mathbf{x'} \mathbf{G}_0(\mathbf{r}-\mathbf{x})\cdot \langle \mathbf{x} | \mathbf{t} | \mathbf{x}' \rangle \cdot \mathbf{G}_0(\mathbf{x}'-\mathbf{r})\\
\nonumber &\approx & \mathbf{G}_{0}(0)  \\
\nonumber &+&  \int d^3\mathbf{x} \int d^3\mathbf{x'} \mathbf{G}_0(\mathbf{r})\cdot \mathrm{e}^{-ik\hat{\mathbf{r}}\cdot \mathbf{x}} \langle \mathbf{x} | \mathbf{t}| \mathbf{x}' \rangle \cdot \mathrm{e}^{-ik\hat{\mathbf{r}}\cdot \mathbf{x}}\mathbf{G}_0(\mathbf{r})\\
\nonumber &=&  \mathbf{G}_{0}(0)  + \mathbf{G}_0(\mathbf{r})\cdot  \mathbf{t}_{k\hat{\mathbf{r}},-k\hat{\mathbf{r}}} \cdot \mathbf{G}_0(\mathbf{r})
\end{eqnarray}
 where we approximated $|\mathbf{r}-\mathbf{x}| \approx r -  \hat{\mathbf{r}}\cdot \mathbf{x} $  valid for $\mathbf{r}$ in the far field and $\mathbf{x}$ typically in the medium, and introduced the $t$-matrix $\mathbf{t}_{kk'}$. The direct Green's function of the Helmholtz equation $\mathbf{G}_0$ far from the object is just the one in free space whose surface integral vanishes trivially. In the far-field the leading term of the backscattered field is purely transverse so that the longitudinal components $\mathbf{EE} +\mathbf{BB}$ of $\mathbf{T}$ defined in Eq.(\ref{TN}) cancel in the momentum leak rate. The latter can thus can thus be written as

 \begin{eqnarray}\label{leakK}
 \nonumber  \mathbf{F}_L \frac{d\omega}{2\pi}=   \frac{2Q\omega}{c_0^2} \frac{d\omega}{2\pi} \, &&\mathrm{Im}  \lim_{r\rightarrow\infty} \frac{\exp(2i(\omega+i\epsilon)r/c_0 )}{(4\pi)^2 }\\
 \nonumber && \times \int d^2 \hat{\mathbf{r}}\, \hat{\mathbf{ r}} \mathrm{Tr} \, (\mathbf{1}-\hat{\mathbf{r}}\hat{\mathbf{r}}) \cdot \mathbf{t}_{k\hat{\mathbf{r}},-k\hat{\mathbf{r}}}(\omega) 
\end{eqnarray}
For any $\epsilon >0$ this leak-induced force vanishes as $r\rightarrow \infty$.  For any finite frequency band, the limit $\epsilon \downarrow 0$ can be done and the oscillating factor  $\exp(2i\omega r/c_0)$ converges ''weakly" to zero as a power law.

In principle the same arguments hold for angular momentum leaks. The angular momentum density of electromagnetic radiation at position $\mathbf{r}$ is given by $\mathbf{r} \times \mathbf{K}$. Conservation of total angular momentum follows directly from Eq.(\ref{mm3})

 \begin{eqnarray}
\nonumber \partial_t \, (\mathbf{r} \times \mathbf{K}) +\partial_t (\mathbf{r}\times \rho \mathbf{v} )= \mathbf{r} \times (\nabla\cdot \mathbf{T})
=  \nabla \cdot ( \mathbf{r }\times \mathbf{T})
\end{eqnarray}
The last equality uses  the symmetry of the tensor $\mathbf{T}$ defined in Eq.~(\ref{TN}) which is crucial to have conservation of angular momentum. Upon integrating over space, the total angular momentum of matter and radiation is

 \begin{eqnarray}\label{consAM}
 \frac{d}{dt}{J}_{\mathrm{tot},i}=  \epsilon_{ijn} \lim_{r\rightarrow \infty } r^3 \int d ^2\hat{\mathbf{r}} \,  \hat{r}_j  T_{nl} \hat{r}_l
\end{eqnarray}
Any contributions to the stress tensor $T_{nl}$ proportional to $\delta_{nl}$ vanish rigourously in this expression, and thus only its components $E_nE_l + B_nB_l$ are relevant, with one field being necessarily along $\hat{\mathbf{r}}$. The free propagator  $\mathbf{G}_0(\mathbf{r})$ is asymptotically transverse to $\hat{\mathbf{r}}$, decaying as $1/r$, the part longitudinal  to $\hat{\mathbf{r}}$  decays at least as $c_0/i\omega r^2$. The factor $r^3 $ is therefore completely compensated and we find for the average couple due to the leak of angular momentum,

  \begin{eqnarray}
\nonumber {N}_{i} \frac{d\omega}{2\pi} \sim \frac{Q }{c_0} \frac{d\omega}{2\pi}\lim_{r\rightarrow \infty }\mathrm{ Re } \frac{\exp(2i(\omega+i\epsilon)r/c_0 )}{(4\pi)^2i } \\
\nonumber \times \int d ^2\hat{\mathbf{r}} \,\epsilon_{ijn} \hat{r}_j t_{n,k\hat{\mathbf{r}},l,-k\hat{\mathbf{r}}}(\omega) \hat{r}_l
\end{eqnarray}
which vanishes again weakly for any finite bandwidth.

\subsection{Angular momentum of a Faraday-active sphere}
We consider a spherical region with volume $V_0$ that exhibits the Faraday effect, i.e. possesses  a dielectric tensor $\varepsilon_{ij} = m^2 (\omega)\delta_{ij} + i (c_0/\omega) V(\omega) \epsilon_{ijk}B_{0,k}$ that is homogeneous inside the sphere, with the Faraday effect quantified by the Verdet constant that measures the amount of rotation per meter per Tesla that would undergo a linearly polarized plane wave in a homogeneous medium. Here, for simplicity, we will assume that $(\varepsilon -1)\omega r/c_0 <1$ and leave the calculation of angular momentum of a genuine Faraday-active Mie sphere to future work. We will demonstrate the existence of a non-zero angular momentum of the radiation, proportional to $Q$ and the magnetic field. Since we consider only  the dielectric tensor, ignore spatial dispersion, with  $\mu=1$ and $\chi=0$, we immediately conclude from the ''no net current law" that $\int d^3\mathbf{r} \langle\mathbf{K}(\mathbf{r})\rangle=0$. Consequently, the average angular momentum $\langle \mathbf{J}\rangle $ \emph{is independent on the origin }with respect to which we calculate the angular moment, quite similar to the magnetic moment of a bounded region of divergenceless currents \cite{jackson}. Because total angular momentum is conserved, any slow change in the magnetic field will make the object start rotating, quite analoguous to the De Haas-Einstein effect\cite{dehaas}.

Following the approach by Ref.\cite{cohenQED} we express the total angular momentum $\mathbf{J}$ of the radiation in Fourier components, and divide it into three contributions: $\mathbf{L}$ (orbital angular momentum), $\mathbf{S}$ (spin) and $\mathbf{J}_c$ associated with the longitudinal electric field and contribution to the ''canonical" angular momentum attributed to the matter,

  \begin{eqnarray}
\mathbf{J}=\int d^3 \mathbf{r} \, \mathbf{r}\times \mathbf{K}(\mathbf{r})   = \mathbf{L}+\mathbf{S}+\mathbf{J}_c
\end{eqnarray}
When applied to the random field satisfying (\ref{GG}) we find for the spectral densities
  \begin{eqnarray} \label{LS}
\nonumber \langle L_i (\omega)\rangle &=& \frac{Q}{-ic_0^2} \epsilon_{ijn} \sum_\mathbf{k} k_n \nabla_j G^\bot_{nn}(\omega,\mathbf{k},\mathbf{k}')_{\mathbf{k}=\mathbf{k}'} + h.c \\
\langle S_i (\omega) \rangle &=& \frac{Q}{-ic_0^2} \epsilon_{ijn} \sum_\mathbf{k}  G^\bot_{jn}(\omega,\mathbf{k},\mathbf{k})  +h.c
\end{eqnarray}
with $\nabla_j = \partial/\partial k_j$  and $\mathbf{G}^\bot(\omega,\mathbf{k},\mathbf{k}') = (1-\hat{\mathbf{k}}\hat{\mathbf{k}})\cdot \mathbf{G}(\omega,\mathbf{k},\mathbf{k}')\cdot (1-\hat{\mathbf{k'}}\hat{\mathbf{k'}})$ is a purely transverse Green's function. The canonical angular momentum is,
  \begin{eqnarray}\label{Jcan}
\langle J_{c,i} (\omega) \rangle &=& \frac{Q}{ic_0^2} \epsilon_{ijn} \sum_\mathbf{k} \nabla_j k_l G^\|_{ln}(\omega,\mathbf{k},\mathbf{k}')_{\mathbf{k}=\mathbf{k}'} + h.c
\end{eqnarray}
where $\mathbf{G}^\| = \hat{\mathbf{k}}\hat{\mathbf{k}} \cdot \mathbf{G}(\omega,\mathbf{k},\mathbf{k}')\cdot (1-\hat{\mathbf{k'}}\hat{\mathbf{k'}})$ is longitudinal on the left and transverse on the right.

The next step is to express the Green's function in terms of the Faraday effect. In terms of the $t$-matrix of the sphere this relation is just
  \begin{eqnarray}
\nonumber \mathbf{G}(\omega,\mathbf{k},\mathbf{k}') = \mathbf{G}_0(\omega, \mathbf{k})\delta_{\mathbf{kk}'} + \mathbf{G}_0(\omega, \mathbf{k}) \cdot \mathbf{t}_{\mathbf{k}\mathbf{k}'}(\omega)\cdot \mathbf{G}_0(\omega, \mathbf{k}')
\end{eqnarray}
Because we neglect multiple Mie scattering in the sphere, the only term in the $t$-matrix that contributes to angular momentum is the Faraday effect inside the sphere
   \begin{eqnarray}\label{tsphere}
 {t}_{i\mathbf{k}j\mathbf{k}'}=  \frac{i\omega V V_0}{c_0} \epsilon_{ijk}B_{0,k} S(\mathbf{k}-\mathbf{k}')
  \end{eqnarray}
  where the $ S(\mathbf{q})  =V_0^{-1} \int_S d^3 \mathbf{x }\exp(i \mathbf{q} \cdot \mathbf{x})$  is the normalized structure function of the sphere. It is straightforward to insert this expression into the one for the angular momentum and to perform the integrals over $\mathbf{k}$. Note that $\nabla_j \exp[i(\mathbf{k}-\mathbf{k}')\cdot \mathbf{r}] _{\mathbf{k}=\mathbf{k}'}  = ir_j$ whose integral over the sphere equals $i\mathbf{r}_0$ with $\mathbf{r}_0$the center  of the sphere. However, the remainder does not survive the odd $k$-integrals in (\ref{LS}) and (\ref{Jcan}) so that the angular momentum becomes  becomes independent on the chosen origin as announced earlier, and proportional to the volume $ V_0$ of the sphere. The $k$-integrals in (\ref{LS}) and (\ref{Jcan})  converge as $d^3\mathbf{k} /k^4$ and we obtain
    \begin{eqnarray}
 \langle \mathbf{L}\rangle=\langle \mathbf{S}\rangle = -V_0\frac{Q(\omega)V(\omega) }{6\pi c_0^2}  \mathbf{B }_0
  \end{eqnarray}
The canonical angular momentum generates a $k$-integral that diverges as $d^3\mathbf{k} /k^2$ at large $k$-vectors but this divergency cancels in the difference $\mathbf{G}(\mathbf{k}) -\mathbf{G}^\dag(\mathbf{k})$. We find,
    \begin{eqnarray}
 \langle \mathbf{J}_c\rangle=  -2V_0\frac{Q(\omega)V(\omega) }{3\pi c_0^2}  \mathbf{B }_0
  \end{eqnarray}
As a result,
    \begin{eqnarray}
 \langle \mathbf{J} \rangle= -V_0\frac{Q(\omega)V(\omega) }{\pi c_0^2}  \mathbf{B }_0
  \end{eqnarray}
In this simple model, the total angular momentum of a Faraday-active region in a sea of isotropic electromagnetic noise is proportional to the applied magnetic field, the volume,  the  power spectrum $Q$ and the bandwidth. For a particle size of $10$ $\mu$m, $VB_0=100$ rad/m, and a noise energy rate $Q \Delta \omega = 10^{-7} $ J/s (corresponding to thermal noise at room temperature or equivalently to a laser with flux $100 \, kW/m^2$ and bandwidth $\Delta \omega = 10^5 $ Hz, this is of order $J\Delta \omega = 30 \hbar$, which is again very small for a macroscopic object.

To see how the angular momentum emerges in real space it is instructive to look at the momentum density $\mathbf{K}(\mathbf{r})$ of the radiation as a function of the distance from the center of the sphere. For isotropic radiation scattering from a magneto-optical sphere, $\mathbf{B}_0$ and $\mathbf{r}$ are the only two vectors relevant for $\mathbf{K}(\mathbf{r})$, which must be of the form
\begin{eqnarray}\label{Krot}
  \langle\mathbf{K}( \mathbf{r},\omega)  \rangle=  Q(\omega) V(\omega) f(\omega, r) \, \mathbf{ r }\times \mathbf{B}_0
\end{eqnarray}
It is easy to see that $\nabla\cdot \langle\mathbf{K} \rangle=0$ and that we can write $\langle \mathbf{K} \rangle = \nabla \times\mathbf{ V}(\mathbf{r}) $, introducing the pseudo-vector field
\begin{eqnarray*}
  \mathbf{V}( \mathbf{r}) =  Q(\omega) V(\omega)  v(\omega, r)\,  \mathbf{r}\times (\mathbf{ r }\times \mathbf{B}_0)
\end{eqnarray*}
provided that $d(vr)/dr = - f$. In terms of $\mathbf{V}(\mathbf{r})$, the total angular momentum is expressed as
\begin{eqnarray*}
 \langle\mathbf{J} \rangle &=& \int d^3 \mathbf{r}\,  \mathbf{r}\mathbf{}\times (\nabla \times\mathbf{ V}(\mathbf{r}))\\
&=& 8 \pi  QV  \mathbf{ B}  \lim_{r\rightarrow \infty} r^5 v(\omega, r)  + 2 \int d^3 \mathbf{r}  \mathbf{ V}(\mathbf{r})
\end{eqnarray*}
The first term will be shown to vanish weakly for any finite frequency interval, the second term is finite and proportional to $\mathbf{B}_0$, identifying $2\mathbf{V}(\mathbf{r})$ as the ``local" density of angular momentum.  If the ``Im $G$" theorem is satisfied we have
\begin{eqnarray*}
  K_n( \mathbf{r}) &=& \frac{Q(\omega)}{ 2 i c_0^2} \left[\partial_n G_{kk}(\mathbf{r},\mathbf{r}') - \partial_j G_{nj}(\mathbf{r},\mathbf{r}') \right]_{\mathbf{r}=\mathbf{r}'} + h.c.
\end{eqnarray*}
In the far field of the sphere, the field scattered from $\mathbf{r}$ to $\mathbf{r}'$  is given by $\delta \mathbf{G} (\mathbf{r},\mathbf{r}') = \mathbf{G}_0(\mathbf{r}) \cdot \mathbf{t}_{k\mathbf{\hat{r}},-k\mathbf{\hat{r}'}}  \cdot  \mathbf{G}_0(\mathbf{r}') $ with the free electromagnetic propagator $\mathbf{G}_0(\omega,\mathbf{r}) = -[ P(y) \Delta_\mathbf{r} + Q(y)  \mathbf{\hat{r}}\mathbf{\hat{r}} ] /4\pi r$ in terms of $y=\omega r/c_0$. With the $t$-matrix given in Eq.~(\ref{tsphere}) we find,
 \begin{eqnarray*}
  f(\omega,r)  &=&  \frac{2 V_0 S(2\omega/c_0)\omega^2 }{ (4\pi)^2 r^3 c_0^4} \mathrm{Im} \, Q(y)  \left[ P'(y) -\frac{Q(y)}{y} \right]
\end{eqnarray*}
Asymptotically $QP' \sim \exp(2iy)/y$ which implies $f(r)\sim  \sin(2\omega r/c_0)/r^4$ and $v(r)\sim  \cos(2\omega r/c_0)/r^5$. As a result, the surface term above vanishes in the weak sense.

\subsection{Finite energy current? }

Equation~(\ref{Krot}) shows the presence of a momentum current circulating around the external magnetic field and should apply for general Mie scattering. It is reminiscent of the contribution of an external magnetic field to the canonical momentum of a charged particle. Its charge is here replaced by the Verdet constant which is charge-odd. The momentum density depends on the radial distance, unlike a charge in a homogeneous magnetic field. This radial dependence extends outside the Faraday-active medium and finally decays as as an oscillating power law. We note that since $\mathbf{B}=\mathbf{H}$,  the Poynting vector  $\mathbf{S}= \mathbf{K} c_0^2$ also circulates around the sphere, satisfying  both $\nabla\cdot \mathbf{S} =0 $ and  the ``no net current" theorem~(\ref{Sint}). The presence of a stationary energy current seems surprising in view of the  isotropic noise that triggers the effect. Textbooks \cite{jackson} say that the energy current density is determined up to a curl of a vector field. This implies that the genuine energy current here could be given by
$c_0 \mathbf{E} \times\mathbf{ H}/4\pi - \nabla \times \mathbf{V } =0$ which would then vanish \emph{everywhere}.  In homogeneous media, this ambiguity is eliminated since Poynting vector $c_0 \mathbf{E} \times\mathbf{ H}/4\pi $ and group velocity $d\omega/d\mathbf{k}$ are  parallel when the material parameters conserve energy \cite{LLgroup}. The presence of an energy current circulating in the near field of the magneto-optical sphere, decaying with distance as $\langle {S}_\phi(r) \rangle\sim  \int d\omega \sin(2\omega r)  /r^3$, i.e. at least as $1/r^4$, has to be investigated without just relying on conservation laws, and needs further attention, for instance by using a time-dependent treatment. The liberty to add the curl of a vector field does not apply for the angular momentum density $\mathbf{r }\times (\mathbf{E} \times\mathbf{ B})/4\pi c_0 $  since it does not occur as a divergence in the conservation law, but rather as a time-derivative.

\section{Conclusions and Outlook}

In this work we have considered an electromagnetic radiation field in anisotropic, and when necessary, bi-anisotropic, conservative  materials. The radiation is assumed to be broad-band and in detailed balance.  In this case it is well-known that the electromagnetic field correlations can be expressed in terms the Green function of the Helmholtz equation. We have established that the space-integrated average Poynting vector vanishes, and only in a homogeneous medium this implies that the Poynting vector vanishes everywhere. In bi-anisotropic media this result is not trivial since opposite wave numbers $\mathbf{k}$ and $-\mathbf{k}$  do not cancel. The momentum density, however, does not vanish in the presence of bi-anisotropic behavior, and one example is a material that exhibits both  optical activity and Faraday rotation. Finally we have considered the angular momentum of the radiation in a simple model with the Faraday effect present in a finite sphere. We find a \emph{nonzero } angular momentum proportional to the magnetic field which implies that any slow change in the magnetic field will exert a torque on the matter. It would be interesting to calculate this effect for a genuine magneto-optical Mie sphere. Unfortunately, these finite momenta are very small in the presence of thermal noise. Our main conclusion is that broken symmetry in matter can lead to radiation forces even if the background radiation is entirely isotropic. Their existence is surprising and a contribution to electrodynamics in general. However, this broken symmetry should be significant over very large bandwidths to make these effects observable.

I would like to thank Geert Rikken for his interest, and the European Space Agency as well as ANR for their support in early stages of this work.


\begin{thebibliography}{References}
\bibitem{nelson} D.F.Nelson, Phys. Rev. A \textbf{44,} 3985 (1991). In this reference a material contribution $t_{il}$ to the stress tensor $\mathbf{T}$ is obtained that is also subject to dynamic dispersion, that is beyond the scope of the present work.
\end{thebibliography}
\end{document}